\documentclass[namedreferences]{solarphysics}

\usepackage[hyperref,optionalrh]{spr-sola-addons} 
\usepackage{graphicx}        
\usepackage{color}           
\usepackage{breakurl}        

\chardef\us=`\_

\begin{document}
\begin{article} 
\begin{opening}
\title{Solar Energetic Particle Events with Protons above 500~MeV between 1995 and 2015 Measured with SOHO/EPHIN}

\author[addressref=aff1,email={kuehl@physik.uni-kiel.de}]{P. K\"uhl}	
\author[addressref=aff1]{N. Dresing}
\author[addressref=aff1]{B. Heber}
\author[addressref=aff1]{A. Klassen}

\address[id=aff1]{Institute for experimental and applied physics, University Kiel, 24118 Kiel, Germany}

\runningauthor{P. K\"uhl \textit{et al.} 2016}
\runningtitle{SEP Events with Protons above 500~MeV}

\begin{abstract}
The Sun is an effective particle accelerator producing solar energetic particle (SEP) events during which particles up to several GeVs can be observed. Those events observed at Earth with the neutron monitor network are called ground level enhancements (GLEs). Although these events with a high energy component have been investigated for several decades, a clear relation between the spectral shape of the SEPs outside the Earth's magnetosphere and the increase in neutron monitor count rate has yet to be established. Hence, an analysis of these events is of interest for the space weather as well as the solar event community.\newline
In this work, SEP events with protons accelerated to above 500~MeV have been identified using data from the \textit{Electron Proton Helium Instrument} (EPHIN) aboard the \textit{Solar and Heliospheric Observatory} (SOHO) between 1995 and 2015. For a statistical analysis, onset times have been determined for the events and the proton energy spectra were derived and fitted with a power law.\newline
As a result, a list of 42 SEP events with protons accelerated to above 500~MeV measured with the EPHIN instrument onboard SOHO is presented. The statistical analysis based on the fitted spectral slopes and absolute intensities is discussed with special emphasis on whether or not an event has been observed as GLE. Furthermore, a correlation between the derived intensity at 500~MeV and the observed increase in neutron monitor count rate has been found for a subset of events.
\end{abstract}


\end{opening}

%

\section{Introduction}
The first solar energetic particle (SEP) event that is now called a ground-level enhancement (GLE) was reported by \cite{Forbush-1946}. GLEs are large SEP events that are observed by ground based experiments such as neutron monitors (NMs). These detectors measure secondary particles produced when ions with energies above several hundred of MeVs create a nuclear cascade in the Earth's atmosphere. Since 1942 71 GLEs have been reported (see \textit{e.g.} https://gle.oulu.fi/) with the largest measured increase above the pre-event background of about 4500\% being observed during GLE 5 on February 23, 1956 \citep{reames-13}. \newline
In order to fully understand the physics behind the particles resulting in GLEs, the chain of acceleration in the corona, the injection and transport in interplanetary space as well as the propagation through the Earth's magnetosphere and atmosphere has to be understood. Therefore \citet{Mishev-etal-2013} calculated the atmospheric yield function that describes the relationship between the intensity of protons and $\alpha$-particles near Earth and the neutron monitor count rate showing significant values for proton energies above 700~MeV. In agreement to these findings, investigations by \cite{Gopalswamy-etal-2014} using GOES measurements of protons with energies above 700 MeV showed a good correlation between the occurrence of above 700 MeV SEPs and GLEs during solar cycle 23 and 24. However in an extended study, \citet{Thakur-2016} have reported two exceptions out of the 16 GLEs during solar cycle 23 and 24. The event of 6 May 1998 caused a GLE but did not cause an increase in the above 700 MeV GOES measurements and the 8 November 2000 that caused an increase of above 100\% regarding the pre-event background in the GOES channel but no increase in the neutron monitor network. \newline
This dilemma where a solar energetic particle event with a proton intensity increase at energies above 700~MeV observed in the near Earth environment is recorded as a GLE and vice versa does also depend on the measurement capabilities of the available instruments. While NMs are a valuable tool to investigate GLEs, they have several limitations due to the indirect nature of detection. By measuring the count rate of secondary particles at ground created by interactions of high energy particles with the atmosphere, NMs do not provide any direct information regarding the interplanetary spectrum of particles. In addition, the Earth's magnetosphere and hence the resulting geomagnetic cutoff rigidity \citep{lockwood-99} can vary over time, further increasing the uncertainties in the analysis of NM data. Therefore, in addition to simulations of these magnetospheric and atmospheric effects, knowledge of the energy spectrum outside of the magnetosphere is required. Recently we showed that the Electron Proton Helium Instrument (EPHIN) is capable to measure proton energy spectra up to 1~GeV \citep{kue15a,kue15b,kue16a,heber_icrc} providing the necessary data for this kind of investigations. \newline
In this work, this new data is used to identify SEP events with protons above 500~MeV during the time period from 1995 to 2015. A detailed comparison with other event lists has been carried out. Furthermore, a statistical analysis of the events based on their spectral properties and of the neutron monitor count rate increase for events resulting in GLEs is presented.\newline
The article is structured as follows, first the instrumentation and data are described prior to the identification of events and the compilation of an event list. Then, a statsistical analysis of the event list by analyzing the proton spectra is presented.


\section{Instrumentation and Data}
\subsection{Validation of High Energy Proton Channels during different Event Phases}
    \begin{figure*}
   \includegraphics[width=0.49\textwidth]{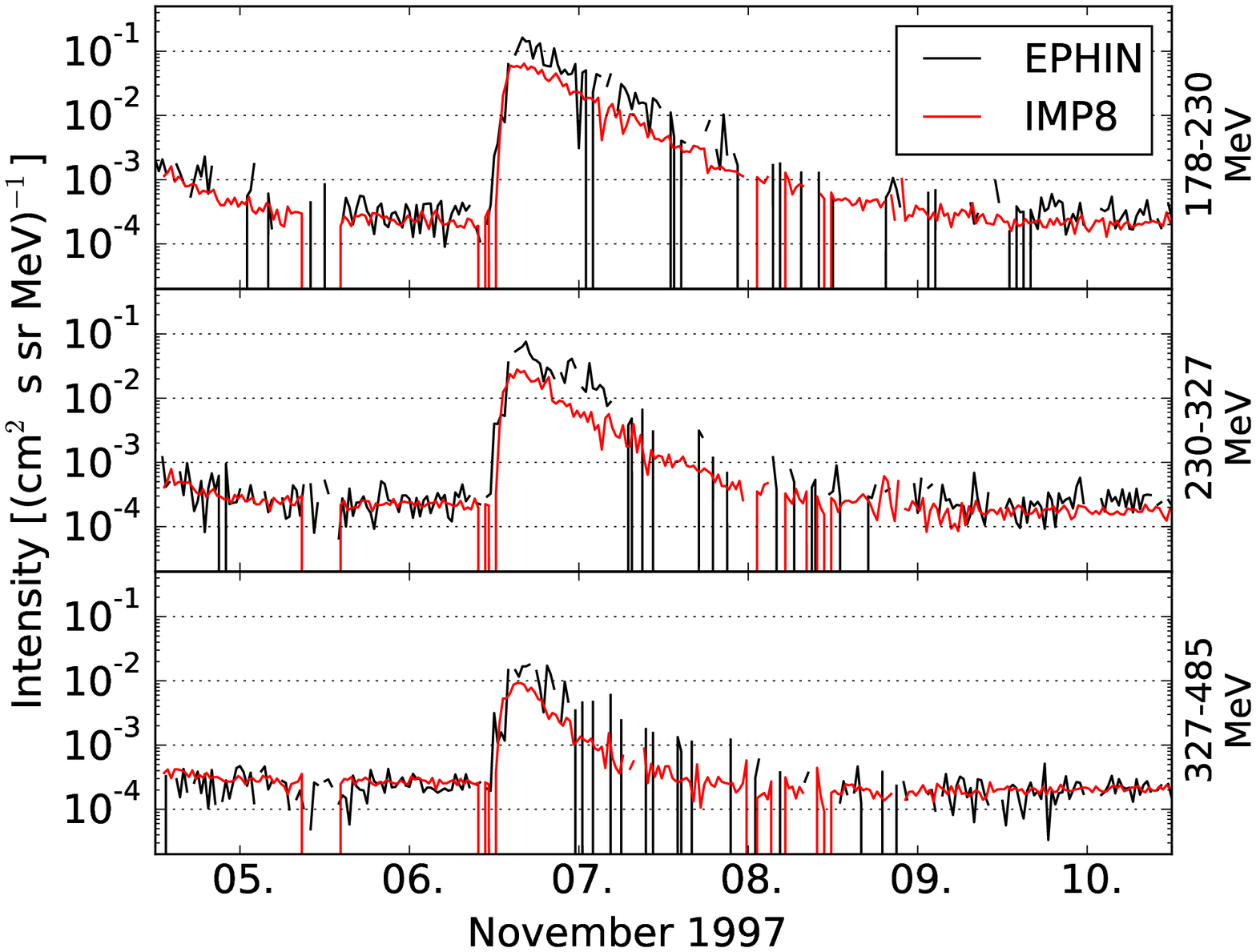}  
   \includegraphics[width=0.49\textwidth]{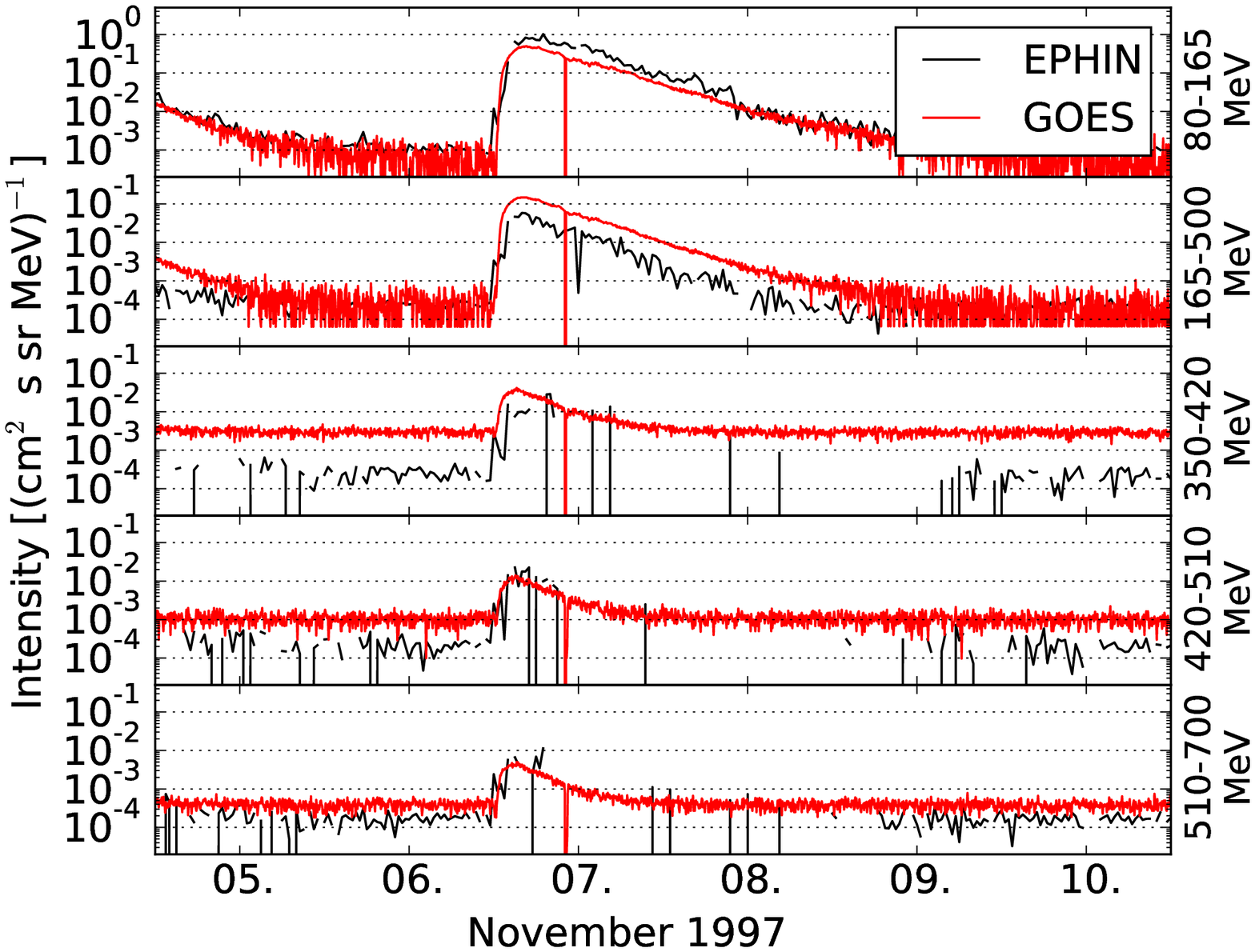}   
   \caption{Proton intensities of the November 1997 GLE measured in different energy channels by IMP-8/GME (left panel, red curve) as well as GOES-9/EPS and GOES-9/HEPAD (right panel, red curve). For comparison, intensities for the same energy channels derived in this study from SOHO/EPHIN data (black curves) are presented.}
              \label{fig:comparison}%
    \end{figure*}   
The EPHIN instrument \citep{mue95} onboard the \textit{Solar and Heliospheric Observatory} (SOHO) consists of a stack of six silicon semiconductors labeled A to F surrounded by an anticoincidence. The nominal energy range between 5 and 50~MeV for protons was extended to energies from 100~MeV to above 1~GeV with a method first presented by \citet{kue15a}. The method is based on particles that penetrate the entire detector stack, depositing only a fraction of their kinetic energy in the instrument. For these penetrating particles, the energy deposition in the detectors C and D are taken as measurement of the energy losses of these particles. It has been shown that for energy losses in a certain range a reliable particle identification is possible for the penetrating particles. The energy losses can then be converted back to total kinetic energy with an uncertainty between 10~\% (at 100~MeV) and 20~\% \citep[at 1~GeV, \textit{c.f.} figure 9 in][]{kue15a}. For a more detailed description of the method, please refer to \citet{kue15a} and \citet{kue16a}.\newline
The method has been succesfully validated for the solar energetic particle events on 17 May 2012 and 6 January 2014 \citep{kue15a}. \citet{kue16a} have proven that the method is also applicable in the absence of solar events to derive galactic cosmic ray (GCR) spectra from 250~MeV up to 1.6~GeV.\newline
In this work, the method is used to identify solar energetic particle events with protons accelerated to at least 500~MeV. To apply this method to the entire SEP event, a further validation of the method during the entire solar event (including the rising and declining phases) is necessary since \citet{kue15a} have only calculated event spectra for certain time periods. Since the method reconstructs the energy individually for every particle detected, it provides the opportunity to define any arbitrary energy channel between $\approx$~100~MeV up to above 1~GeV. Hence, intercalibration and comparison with other missions can be achieved rather easily.\newline 
For this purpose, Figure \ref{fig:comparison} shows the intensity of different energy channels during the 6 November 1997 SEP event (GLE~55) measured by the \textit{Goddard Medium Energy Experiment} \citep[GME:][]{imp_gme} onboard \textit{Interplanetary Monitoring Platform 8} (IMP-8, left) as well as by the \textit{Energetic Particle Sensor} \citep[EPS:][]{goes_eps} and \textit{High Energy Proton and Alpha Detector} \citep[HEPAD:][]{goes} onboard GOES-9 (right). In addition, the measured SOHO/EPHIN intensities in the same energy range are shown (black). From the figure, it is evident that all three instruments measure the SEP event and that the intensity-time-profiles are in agreement within a factor of two. \newline
However, EPHIN measures systematical higher intensities around the maximum and in the decay phase of the event when compared to the IMP-8 instrument in all three channels. In contrast, the intensities are in agreement before and after the event. Hence we attribute the differences during the event to the so-called ring switching \citep[for details see][]{mue95}. \newline
It is important to note that the pre-event background measured by GOES is an order of magnitude higher than the ones given by EPHIN and the IMP-8 instrument as already described by \citet{goes_imp}. Furthermore, the 30 minutes averaged data from SOHO/EPHIN have statistical limitations, especially in the decay phase of the event.\newline

\subsection{Electron Contamination of the High Energy Proton Measurements}
   \begin{figure}
   \centering
   \includegraphics[width=0.49\textwidth]{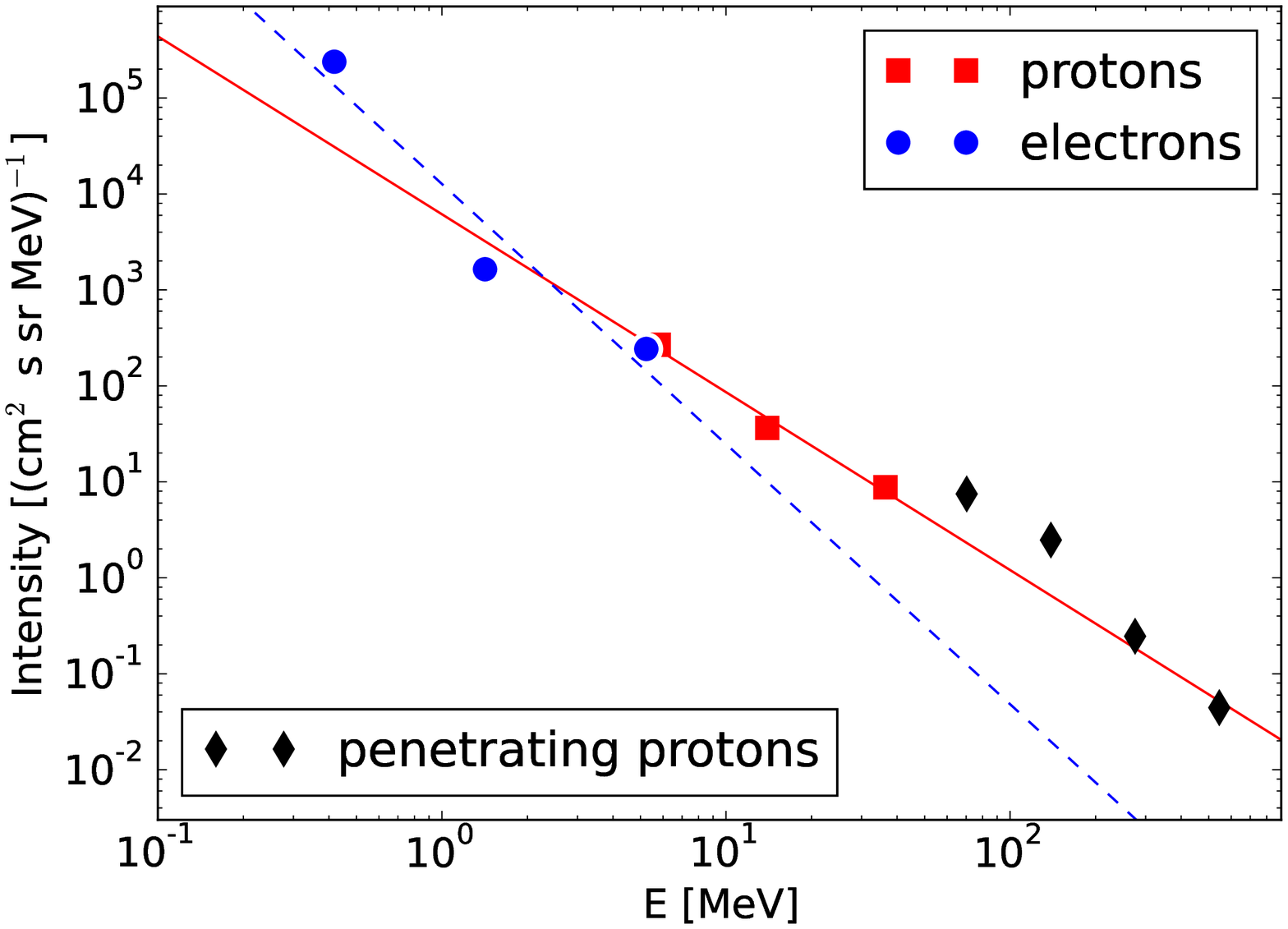}	
   \includegraphics[width=0.49\textwidth]{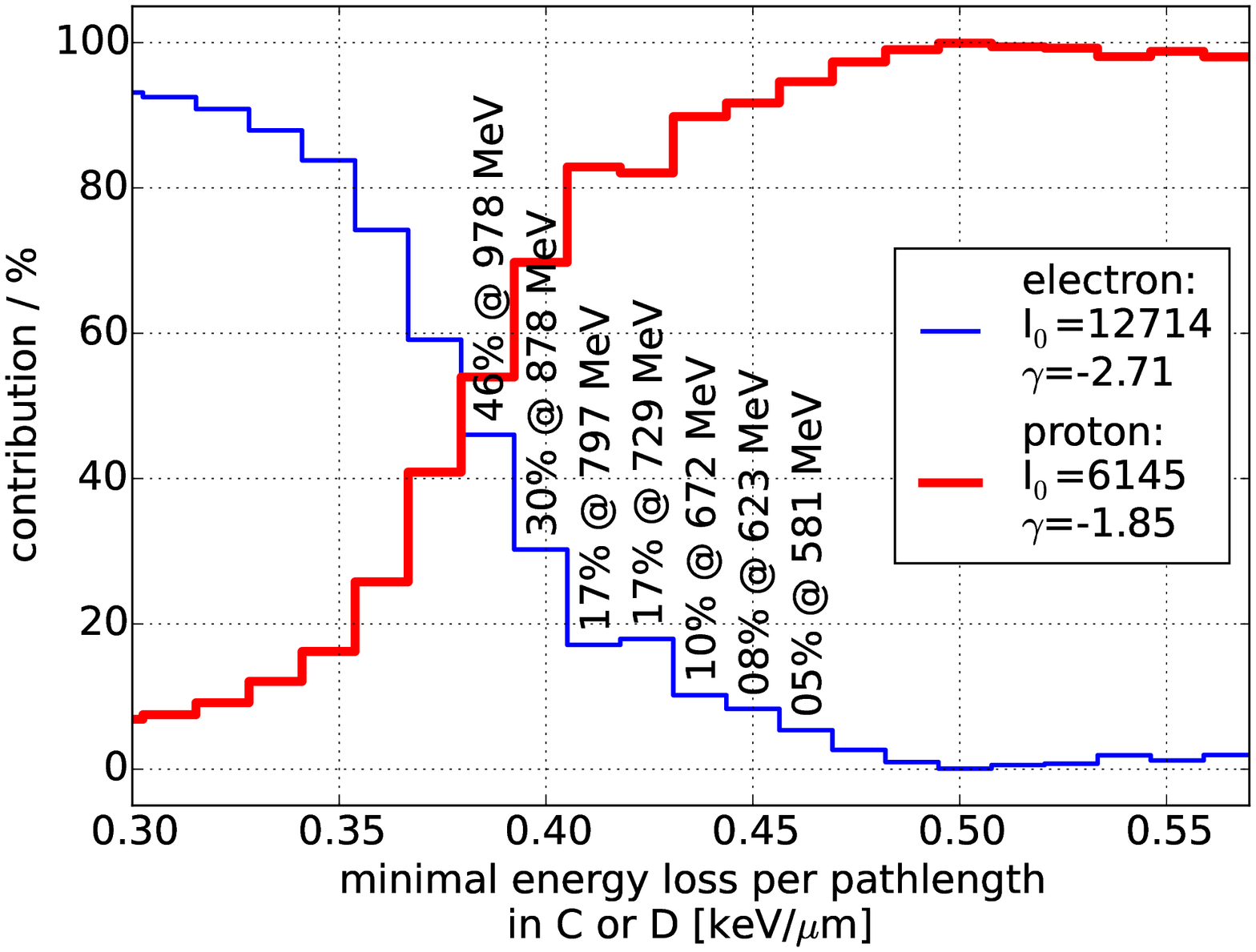}    
   \caption{Left: Proton (red squares) and electron (blue circles) spectra from the nominal data products as well as power law fits (lines) for GLE~69 on 20 January 2005 from 0900\,--\,1100UT. For the same time period, the spectra of protons penetrating the instrument has been derived as well (black diamonds). Right: Simulation results showing the contribution of protons and electrons for different energy losses based on the extrapolated spectra. The extrapolation has been done with a power law fit for both particle spectra with the fit results given in the legend. For some exemplary energy losses, the electron contribution as well as the proton energy related to that energy loss are shown by the text included in the vertical direction.}
              \label{fig:electron}%
    \end{figure}  
Though \citet{kue15a} have mentioned a possible influence of electron fluxes above 10~MeV on the high energy proton data during solar events, this issue has not been quantified yet. 
\citet{heber_icrc} showed that there are discrepancies in the derived proton spectrum during GLE~69 between SOHO and neutron monitors above 700~MeV while GOES and SOHO agree with each other at lower energies. The increased flux above 700~MeV is believed to be caused by electrons above 10~MeV associated to the same SEP event that cause similar energy losses in the detector compared to those of high energy protons. Therefore, the influence of electrons on the derived proton flux above 700~MeV has to be taken into account and a study of this effect is necessary before a detailed analysis of the spectral properties can be performed.\newline
For that purpose, Figure \ref{fig:electron} (left panel) presents electron (blue circles) and proton (red squares) spectra based on the nominal data products from the EPHIN instrument during the GLE~69 on 20 January 2005 from 0900\,--\,1100UT. The event has one of the highest electron contributions among those events investigated in this study and can therefore be considered as the worst case scenario. The spectra have been fitted with a power law and have been extrapolated to higher energies. Note that although the derived proton spectra based on the penetrating particles (black diamonds) are in agreement with the power law fitted to the proton spectrum below 50 MeV, a softening in the proton spectrum at higher energies (\textit{e.g.} a double power law) would increase the electron contribution. Using the fitted spectra as input for a \textit{Geometry and Tracking 4} (GEANT4) monte carlo simulation \citep{geant4} of the instrument, the contribution of both protons and electrons to energy losses in the C and D detector have been derived. As a result Figure \ref{fig:electron} (right) shows the contribution of electrons (blue) and protons (red) dependent on the energy loss. For some illustrative energy losses, the electron contribution as well as the proton energy related to that energy loss are shown by the text included in the vertical direction. From the figure, it can be concluded that the electron contribution to the high energy proton spectra is: 1) negligible below $\approx$500~MeV, 2) less than 20~\% in the energy range from 500 to 800~MeV, and 3) a major contribution above 800~MeV. Hence, proton intensities above 800~MeV should be considered as upper limits during solar events.\newline

\section{Identification Method of $>$500~MeV Proton Events}
\subsection{Event Detection}
\begin{figure}
   \centering
   \includegraphics[width=0.75\textwidth]{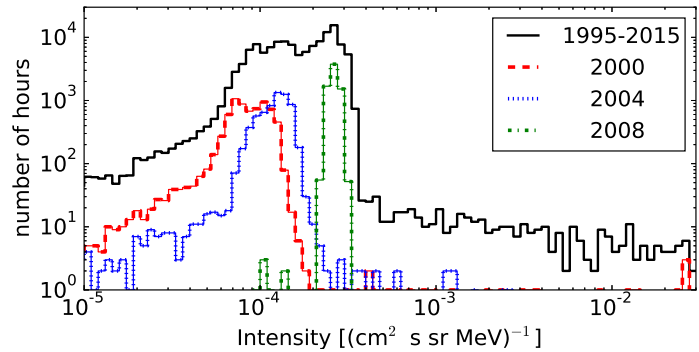}   
   \caption{Histograms of the hourly proton intensities in the energy range from 500 to 700~MeV based on SOHO/EPHIN data. Shown are a histogram of the entire mission, as well as three annual histograms.}
              \label{fig:th}%
    \end{figure}

In order to identify SEPs with protons accelerated to energies above 500~MeV, a histogram of hourly intensities in a defined 500\,--\,700~MeV range from 1995 to 2015 is presented in Figure \ref{fig:th}. The histogram indicates that most of the time, the measured intensity is in the range of $0.7\ - \ 4\cdot10^{-4}$ (cm$^2$ s sr MeV)$^{-1}$. In agreement to \citet{kue16a}, these intensities correspond to the GCR background. The variation of the peak position over different years as indicated by annual histograms can be explained by solar modulation \citep{heber2006a,heber2006b}. While intensities below this main population correspond to either GCR depressions during the passages of interplanetary coronal mass ejections \citep[forbush decreases,][]{cane} or instrumental effects such as a high deadtime of the electronics, higher intensities are related to SEPs.\newline
In this study, events have been identified by requiring that at least two hourly averaged intensities in a six-hour interval are above a threshold of $4\cdot10^{-4}$ (cm$^2$ s sr MeV)$^{-1}$. Using this identification technique, 42 solar particle events in the time between start of the mission (December 1995) and 1 October 2015 have been identified. It has to be noted that communication with SOHO was lost for several months during 1998 and, hence, no EPHIN data is available for this time period. \newline
\subsection{The Event List}
   \begin{figure}
   \centering
   \includegraphics[width=0.75\textwidth]{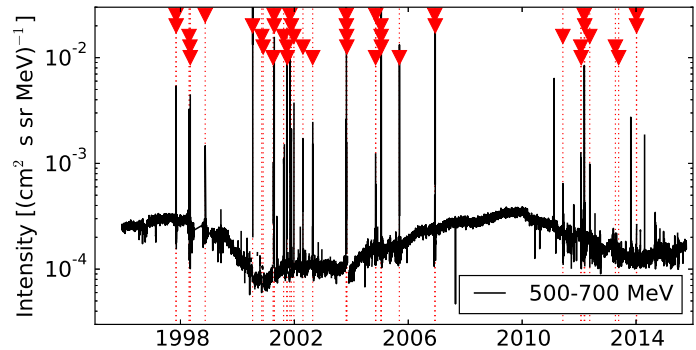}   
   \caption{Time profile of the proton intensity in the energy range from 500 to 700~MeV averaged over six hours for the last 20 years. The dashed, vertical lines and the arrows indicate the dates of the SEP events found in this study.}
              \label{fig:dates}%
    \end{figure}  
    
Figure \ref{fig:dates} presents the time profile of the proton intensity in the energy range from 500 to 700~MeV over the last 20 years. In agreement to Figure \ref{fig:th}, the variation of the GCR background intensity lies between 0.7 and 4~$\cdot10^{-4}$(cm$^2$~s~sr~MeV)$^{-1}$ over Solar Cycles 23 and 24. The dates of the events identified in this study are marked as red triangles. Note that some intensity increases shown in the figure were not selected as events since they were caused by photons from the flare (extreme-ultraviolet to hard X-rays range) depositing energy in the detector stack. Since these photon peaks are usually short lived they do not increase the intensity over two hours and are therefore not identified by the algorithm. From the 42 events, 32 events occured during Solar Cycle 23 and only ten in Solar Cycle 24. As expected, the occurence of the solar events is clearly more likely during solar maxima \citep[around 2002 and 2014,][]{nymmik}.\newline
The dates and times when the events passed the threshold are listed in columns two and three of Table \ref{tab:events}. Note that these numbers mark the time when the event was identified by the algorithm and they should not be confused with onset times. The onset times are derived in the next section. Columns five to eight give the corresponding events from other studies, namely GLEs (taken from \url{http://gle.oulu.fi/}), SEPServer \citep{sepserver}, GOES (major sep list, \url{http://cdaw.gsfc.nasa.gov/CME_list/sepe/}), and \citet{gopalswamy}, respectively. The exact times of the GLEs were taken from the \textit{Neutron Monitor Database} (NMDB, \url{http://www.nmdb.eu/nest/gle_list.php}). The GLE~68 is listed at 00:00 UT by the NMDB while \citet{cliver} relates the GLE to a flare peaking at 09:52 UT. Hence, for this event, the time from \citet{cliver} was adopted.\newline
The events N$^0$35, N$^0$36 and N$^0$41 are listed as Sub-GLEs in the Oulu GLE database (\url{https://gle.oulu.fi/}) and the event N$^0$42 was suggested to be a GLE by \citet{thakur_gle72}. However, they are not officially confirmed to be GLEs and therefore they are not marked as GLEs in our table.\newline
From the table, it is evident that the 42 events include all GLEs during the SOHO age (GLE~55 to GLE~72) with exception of GLE~58, during which SOHO had a data gap. Since GLEs are known to be caused by events during which particles are accelerated to above 500~MeV \citep{cliver83,plainaki,shen13}, this is a validation of the event identification method.\newline
Regarding the fact, that the SEPServer list is based on data from 1996 to 2010, it does also include the majority of the events found in this study. The GOES list features every single event found in this study. \citet{gopalswamy} have derived a list of 37 large solar events for Solar Cycle 24 that occurred until the end of 2014. Only ten of those events were detected by our method, suggesting that the other 27 events did not accelerate protons to energies above 500~MeV. In their analysis, \citet{gopalswamy} identified eight GLE candidates from their list. From those eight events, only three are found to have increased fluxes above 500~MeV based on this study (events N$^0$33, N$^0$38 and N$^0$40).

    \begin{table*}
\begin{center}
\begin{tabular}{ c | c c c c c c c c c }
\hline
N$^0$ & Date & Time & Onset & GLE N$^0$ & SEPS  & GOES  & Gopalswamy  \\  
\hline
01 & 1997-11-04 & 15:00 & 06:00 & - & 2 & 1 & -  \\ 
02 & 1997-11-06 & 13:00 & 12:45 & 55 & 3 & 2 & -  \\ 
03 & 1998-04-20 & 20:00 & - & - & 6 & 3 & -  \\ 
04 & 1998-05-02 & 18:00 & 15:00 & 56 & 7 & 4 & -  \\ 
05 & 1998-05-06 & 13:00 & 08:30 & 57 & 8 & 5 & -  \\ 
06 & 1998-11-14 & 09:00 & 06:15 & - & 12 & 11 & -  \\ 
07 & 2000-07-14 & 12:00 & 10:30 & 59 & 32 & 21 & -  \\ 
08 & 2000-11-09 & 01:00 & 23:30p & - & 38 & 28 & -  \\ 
09 & 2000-11-26 & 18:00 & - & - & - & 30 & - \\ 
10 & 2001-04-03 & 04:00 & 01:00 & - & 45 & 33 & -  \\ 
11 & 2001-04-15 & 16:00 & 15:00 & 60 & 49 & 36 & -  \\ 
12 & 2001-04-18 & 04:00 & 02:45 & 61 & 50 & 37 & -  \\ 
13 & 2001-08-16 & 03:00 & 00:30 & - & - & 42 & -  \\ 
14 & 2001-09-24 & 23:00 & 14:00 & - & 58 & 44 & -  \\ 
15 & 2001-10-01 & 22:00 & - & - & 59 & 45 & - \\ 
16 & 2001-11-04 & 17:00 & 16:30 & 62 & 63 & 48 & -  \\ 
17 & 2001-11-23 & 05:00 & 21:00p & - & 64 & 51 & -  \\ 
18 & 2001-12-26 & 07:00 & 06:30 & 63 & 65 & 52 & -  \\ 
19 & 2002-04-21 & 06:00 & 02:00 & - & 72 & 63 & -  \\ 
20 & 2002-08-24 & 04:00 & 02:15 & 64 & 80 & 71 & -  \\ 
21 & 2003-10-28 & 13:00 & 14:00 & 65 & 88 & 78 & -  \\ 
22 & 2003-10-29 & 22:00 & 21:15 & 66 & - & 79 & -  \\ 
23 & 2003-11-02 & 20:00 & 17:15 & 67 & 90 & 81 & -  \\ 
24 & 2003-11-05 & 06:00 & 02:00 & - & - & 82 & -  \\ 
25 & 2004-11-07 & 21:00 & 15:45 & - & 97 & 90 & - \\ 
26 & 2004-11-10 & 11:00 & 03:15 & - & 99 & 92 & -  \\ 
27 & 2005-01-16 & 14:00 & - & - & 101 & 93 & -  \\ 
28 & 2005-01-17 & 16:00 & 13:45 & 68 & - & 94 & -  \\ 
29 & 2005-01-20 & 09:00 & 06:45 & 69 & - & 95 & -  \\ 
30 & 2005-09-08 & 20:00 & - & - & - & 102 & -  \\ 
31 & 2006-12-06 & 23:00 & - & - & - & 104 & -  \\ 
32 & 2006-12-13 & 04:00 & 03:00 & 70 & 112 & 105 & - \\ 
33 & 2011-06-07 & 09:00 & 07:15 & - & - & 110 & 4  \\ 
34 & 2012-01-23 & 12:00 & 05:30 & - & - & 115 & 9  \\ 
35 & 2012-01-28 & 04:00 & 18:30p & - & - & 116 & 10  \\ 
36 & 2012-03-07 & 04:00 & - & - & - & 117 & 11 \\ 
37 & 2012-03-13 & 19:00 & 17:45 & - & - & 118 & 12  \\ 
38 & 2012-05-17 & 03:00 & 01:45 & 71 & - & 119 & 13  \\ 
39 & 2013-04-11 & 12:00 & 07:45 & - & - & 131 & 25 \\ 
40 & 2013-05-22 & 20:00 & 14:00 & - & - & 133 & 27  \\ 
41 & 2014-01-06 & 10:00 & - & - & - & 137 & 31  \\ 
42 & 2014-01-08 & 00:00 & 21:00p & - & - & 138 & 32  \\ 
\hline
\end{tabular}
\caption{Event list compiled in this article. The columns represent the event number (column 1), the date (column 2) and time (column 3) when the intensity threshold was surpassed, the onset time (column 4), corresponding event numbers of the GLE (\protect\url{http://gle.oulu.fi/}, Sub-GLE are also marked) (column 5), SEPServer \citep{sepserver} (column 6), GOES (\protect\url{http://cdaw.gsfc.nasa.gov/CME_list/sepe/}) (column 7) and \citet{gopalswamy}(column 8) event lists. A \textit{p} in the onset time indicates that the time corresponds to the day prior to the date given in column 1. For details see text.}
\label{tab:events}
\end{center}
\end{table*}
    
\section{Statistical Event Analysis} 
\subsection{Onset Times}
      \begin{figure}
   \centering
   \includegraphics[width=0.7\textwidth]{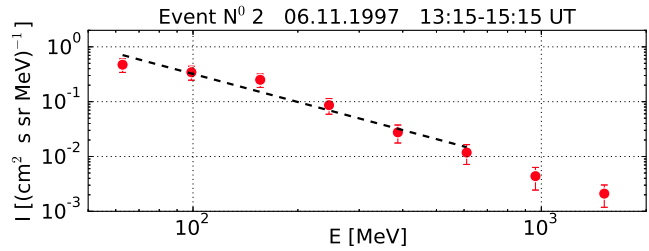}   
   \caption{Proton spectrum with power law fit for event N$^0$2 (GLE 55) from 13:15 to 15:15~UT. Note that only energies below 800~MeV have been used for the fit due to possible electron contamination at higher energies.}
              \label{fig:spectra}%
    \end{figure}  
    
\begin{table*}
\begin{center}
\begin{tabular}{c | c c c c c c c c  }
\hline
lower bin edge [MeV] & 49 & 78 & 124 & 195 & 308 & 486 & 766 & 1207 \\ 
upper bin edge [MeV] & 78 & 124 & 195 & 308 & 486 & 766 & 1207 & 1903 \\ 
geometric mean [MeV] & 62 & 98 & 155 & 245 & 387 & 610 & 962 & 1516 \\ 
\hline
    \end{tabular}
\caption{Energy bins used for the spectra.}
\label{tab:bins}
\end{center}
\end{table*}
For a study of the spectral properties of the events, only those events, for which an onset time based on a 100 to 1000~MeV proton channel could be derived, were taken into account. The chosen broader energy interval in comparison to the 500 to 700~MeV interval allows onset determination with a time resolution of 15~minutes. For 34 of the 42 events it was possible to derive the onset time by requiring an intensity increase above a threshold defined as the average intensity during the previous six hours plus three times the standard deviation of that time interval. These 34 events include all GLEs and Sub-GLEs except for event N$^0$41 since EPHIN had a datagap during the early stage of the event \citep{kue15a}. Hence, no onset could be determined for this event and it is not included in the following analysis. All determined onset times are listed in column four of Table \ref{tab:events}.\newline
\subsection{Event Spectra}
Since the statistics of EPHIN data for the event N$^0$25 are limited, it was excluded from the following study, although an onset time was derived. For the remaining 33 events, the high energy proton spectrum has been calculated in a time interval of two hours starting 30~minutes after the onset. The time lag of 30~minutes is necessary in order to reflect the different travel times of 100~MeV and 1000~Mev protons. While the latter can reach the spacecraft after roughly ten minutes (depending on the length of the Parker spiral and the diffusion in the interplanetary medium), the 100~MeV protons can be delayed by up to half an hour due to their lower velocity. The interval length of two hours was selected for statistical reasons. As an example, Figure \ref{fig:spectra} shows the derived spectrum for event N$^0$2 (GLE~55). In the figure, the geometric mean of the energy range is shown. The exact energy bins are given in Table \ref{tab:bins}.\newline
\citet{mewaldt_gamma} have shown that the proton spectra of GLE events can be reproduced by a double-power law \citep[described by][]{band} with a spectral break at several MeV. Since we only analyze energies above 100~MeV, a single power law function
\begin{equation}
I(E)=I_0\cdot (E/E_0)^\gamma,
\label{eq:powerlaw}
\end{equation} 
where $I(E)$ and $I_0$ are measured in [(cm$^2$ s sr MeV)$^{-1}$] and $E$, $E_0$ in [MeV], has been fitted for every single event. Based on the approximation of the electron contribution described above (\textit{c.f.} Figure \ref{fig:electron}), only energies below 800~MeV have been taken into account for the fit due to possible electron contamination at higher energies. As an example, the fit for event N$^0$2 is shown in Figure \ref{fig:spectra}.\newline
Table \ref{tab:fit} presents the spectral indices [$\gamma$] and intensities at 500~MeV [I$_{500}$] resulting from the fits of the 33 events. The fit was performed as a linear regression between the logarithm of the energy and the logarithm of the intensity. Hence, the goodness of the fits can be represented by the adjusted coefficient of determination [R$^2$] which is also given in Table \ref{tab:fit}. \newline

\begin{table*}
\begin{tabular}{c | c c c || c | c c c}
N$^0$ & $\gamma$ & [I$_{500}$] & \textit{R$^2$} & N$^0$ & $\gamma$ & [I$_{500}$] & \textit{R$^2$}  \\  
\hline
01 & -2.20$\pm$0.16 & (8.77$\pm$1.00)~e-4 & 0.97 & 21 & -3.15$\pm$0.25 & (2.07$\pm$0.15)~e-2 & 0.97 \\ 
02 & -1.70$\pm$0.24 & (2.08$\pm$0.33)~e-2 & 0.95 & 22 & -2.49$\pm$0.11 & (3.70$\pm$0.15)~e-2 & 1.00 \\ 
04 & -2.12$\pm$0.15 & (2.31$\pm$0.21)~e-3 & 0.96 & 23 & -2.70$\pm$0.22 & (1.50$\pm$0.11)~e-2 & 0.98 \\ 
05 & -2.84$\pm$0.26 & (1.15$\pm$0.12)~e-3 & 0.93 & 24 & -2.67$\pm$0.17 & (3.45$\pm$0.30)~e-4 & 0.98 \\ 
06 & -2.44$\pm$0.24 & (1.52$\pm$0.18)~e-3 & 0.91 & 26 & -1.51$\pm$0.11 & (9.92$\pm$1.94)~e-4 & 0.97 \\ 
07 & -2.24$\pm$0.15 & (1.74$\pm$0.11)~e-1 & 0.97 & 28 & -3.20$\pm$0.08 & (6.12$\pm$0.14)~e-3 & 0.93 \\ 
08 & -2.78$\pm$0.31 & (1.32$\pm$0.14)~e-1 & 0.97 & 29 & -2.12$\pm$0.13 & (2.53$\pm$0.14)~e-1 & 0.99 \\ 
10 & -3.24$\pm$0.12 & (5.59$\pm$0.23)~e-4 & 0.98 & 32 & -1.95$\pm$0.15 & (3.56$\pm$0.26)~e-2 & 0.98 \\ 
11 & -2.13$\pm$0.03 & (7.77$\pm$0.10)~e-2 & 0.99 & 33 & -1.83$\pm$0.31 & (9.85$\pm$3.15)~e-4 & 0.85 \\ 
12 & -2.01$\pm$0.12 & (3.79$\pm$0.31)~e-3 & 0.97 & 34 & -3.78$\pm$0.40 & (2.45$\pm$0.27)~e-4 & 0.78 \\ 
13 & -2.57$\pm$0.09 & (2.38$\pm$0.10)~e-3 & 0.99 & 35 & -2.22$\pm$0.24 & (1.65$\pm$0.25)~e-3 & 0.93 \\ 
14 & -3.30$\pm$0.33 & (4.39$\pm$0.47)~e-4 & 0.93 & 37 & -2.75$\pm$0.25 & (1.84$\pm$0.18)~e-3 & 0.94 \\ 
16 & -2.90$\pm$0.20 & (1.06$\pm$0.07)~e-2 & 0.98 & 38 & -1.87$\pm$0.17 & (4.47$\pm$0.57)~e-3 & 0.97 \\ 
17 & -2.22$\pm$0.05 & (5.55$\pm$0.21)~e-4 & 1.00 & 39 & -2.19$\pm$0.09 & (3.40$\pm$0.26)~e-4 & 0.99 \\ 
18 & -3.21$\pm$0.07 & (5.39$\pm$0.11)~e-3 & 0.99 & 40 & -2.93$\pm$0.19 & (5.55$\pm$0.43)~e-4 & 0.92 \\ 
19 & -3.84$\pm$0.10 & (2.35$\pm$0.06)~e-3 & 0.97 & 42 & -3.60$\pm$0.03 & (4.93$\pm$0.04)~e-4 & 1.00 \\ 
20 & -2.58$\pm$0.08 & (4.41$\pm$0.15)~e-3 & 1.00 & \\ 
\end{tabular}
\caption{Results of the power law fit to the proton spectra. The columns represent the event number (\textit{c.f.} Table \ref{tab:events}, column 1), the fitted spectral indices [$\gamma$] (column 2) and intensities at 500~MeV [I$_{500}$] (in units of [(cm$^2$ s sr MeV)$^{-1}$], column 3) of the events. Since the fit was performed as a linear regression between the logarithm of the energy and the logarithm of the intensity, the goodness of the fits is represented by the adjusted coefficient of determination [R$^2$] (column 4). For details see text.}
\label{tab:fit}
\end{table*}

\subsection{Statistical Analysis of $>$500~MeV Proton Events}
   \begin{figure}
   \centering
   \includegraphics[width=0.7\textwidth]{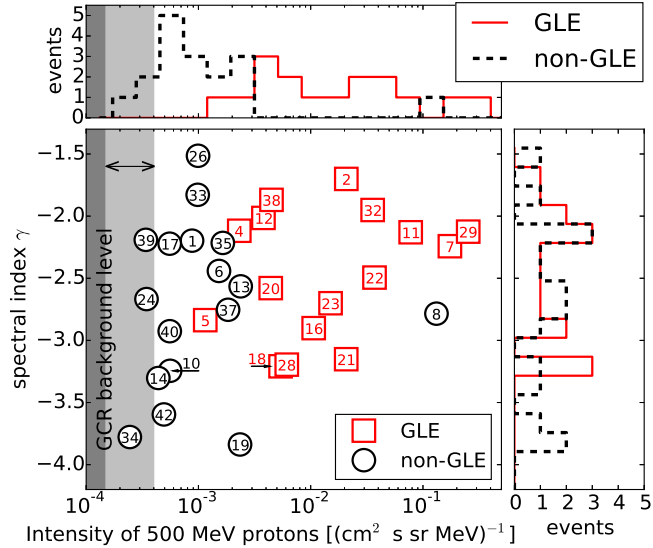}   
   \caption{Bottom left: Spectral index [$\gamma$] as function of intensity at 500~MeV derived from the proton spectra fit. GLEs are marked as red squares, other events as black circles. All numbers correspond to thosen in Tables \ref{tab:events} and \ref{tab:fit}. The grey shaded area marks the GCR background level in terms of intensity. Top and right: Histograms of the intensity at 500~MeV and the spectral index [$\gamma$], respectively. The solid red line corresponds to GLEs. Other events are represented by a dashed black line.}
              \label{fig:fitparas}%
    \end{figure}  
In Figure \ref{fig:fitparas} the spectral index of the analyzed events is shown as a function of the proton intensity at 500~MeV derived from the proton spectra fit. GLEs are shown as red squares, the remaining events as black circles. The numbers in the symbols correspond to those in Tables \ref{tab:events} and \ref{tab:fit}. The dark grey and light grey shaded area correspond to the varying GCR background level at 500~MeV during solar maximum and solar minimum, respectively (\textit{c.f.} Figure \ref{fig:th}). At the top and on the right hand side of the figure, histograms of both quantities are also shown individually.\newline
The fit results of events N$^0$24, N$^0$34 and N$^0$39 show that the proton intensity at the energy of 500~MeV is slightly lower than the threshold used for the event identification based on the 500\,--\,700~MeV channel. This can be explained by statistical errors of both, the channel intensity and the fit results. However, it should be noted that these events occured during solar maximum and hence these events could still have caused an increase above the GCR background. \newline
Event N$^0$08 shows a much higher intensity as well as a similar spectral index compared to several GLEs and, yet, does not show any increases in the neutron monitor count rates which is in agreement with to findings of \citet{Thakur-2016}. Hence, this event is of special interest in terms of understanding which physical processes determine whether or not a SEP event with a certain spectral shape is observed by the neutron monitor network. Therefore, an extensive study using not only the spectral data at high energies but also simulations of the asymptotic viewing directions of neutron monitors is in preparation. \newline
The majority of the GLEs feature spectral indices uniformly distributed between $-2$ and $-3$, which is in good agreement with the results from \citet{mewaldt_gamma}. The spectral indices of events not related to GLEs are also uniformly distributed, but in a wider range between $-2$ and $-4$. The spectral index of event N$^0$20 ($\gamma$=-2.58) is identical with the findings to \citet{tylka06}.\newline
   \begin{figure}
   \centering
   \includegraphics[width=0.7\textwidth]{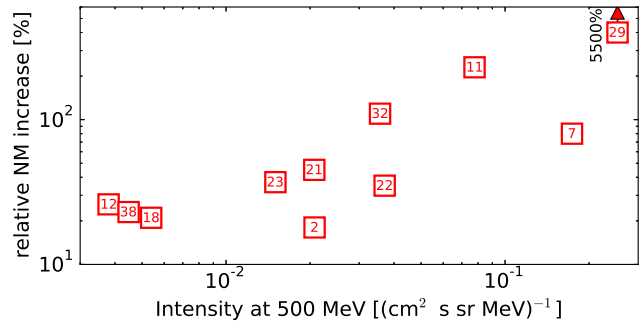}   
   \caption{Relative increase of the neutron monitor count rate as function of intensity at 500~MeV based on the fit. The increase in the NMs are taken from \citet{mccracken_2012}. The numbers correspond to Tables \ref{tab:events} and \ref{tab:fit}.}
              \label{fig:i500_nm}%
    \end{figure}  
However, the intensity of 500~MeV proton is typically higher during GLE related events than during events without GLEs. Furthermore, this intensity is above $2\cdot 10^{-3}$(cm$^2$ s sr MeV)$^{-1}$ for all GLEs (except for event N$^0$5), which is consistent with results from \citet{nitta}. In their study, they showed that the GOES HEPAD P9 channel (420\,--\,510~MeV) usually exceeds $2\cdot 10^{-3}$(cm$^2$ s sr MeV)$^{-1}$ during GLEs with only GLE~57 and GLE~68, corresponding to events N$^0$05 and N$^0$28 in our list, having lower fluxes. \citet{Thakur-2016} also found GLE~57 to be an especially small event. While the results of this work do confirm the results of previous studies regarding the small intensity of GLE~57, the intensity of GLE~68 is found to be higher compared to the results of \citet{nitta}. However, in their study \cite{nitta} noted that the onset determination and hence the analysis of this particular event is troublesome.\newline
The five GLEs with the highest flux in our analysis (events N$^0$29, N$^0$07, N$^0$11, N$^0$22, N$^0$32 corresponding to GLEs 69, 59, 60, 66, 70) are also considered to be among the largest GLEs in the SOHO era \citep[\textit{c.f.} Table 1 in][]{mccracken_2012}. The relative increases in neutron monitor count rate for GLEs given by \citet{mccracken_2012} are shown as function of the intensity at 500~MeV derived in this study in Figure \ref{fig:i500_nm}. Some GLEs from Table \ref{tab:events} are not shown here, as they have not been investigated by \citet{mccracken_2012}. The figure indicates a correlation between the intensity at 500~MeV and the relative increase in NM count rate with the exception of the events with intensities below  $10^{-2}$(cm$^2$ s sr MeV)$^{-1}$ (Events N$^0$12, N$^0$18 and N$^0$38) as well as event N$^0$29, which has a significant higher increase in NM count rate compared to the measured intensity at 500~MeV. Furthermore, it has to be noted that the scattering of the relative NM increase for a given intensity at 500~MeV is rather large. The reasons for these deviations remain unclear especially since the spectral shape of the events (e.g. the fitted $\gamma$) are rather similar for most of them (\textit{c.f.} Figure \ref{fig:fitparas}). Possible explanations are the asymptotic viewing direction of each neutron monitor \citep{mccracken62,smart2000} or changes in the cutoff rigidities due to geomagnetic disturbances during the SEP events \citep{danilova99} which may differ from event to event.\newline

\section{Summary}
In this study, SEP events with protons with energies above 500~MeV have been investigated based on the extended measurement range of SOHO/EPHIN described by \citet{kue15a} and \citet{kue16a}.\newline
We have shown that the new and unique data product is valid during any stage of the solar event (e.g. onset, maximum and decay phase) by comparison with results from IMP8 and GOES. Furthermore, additional simulations of the instrument have shown that the electron contribution to the high energy proton data is: 1) negligible below 500~MeV, 2) less than 20\% between 500 and 800~MeV, and 3) seriously uncertain above 800~MeV.\newline
Using the energy interval from 500 to 700~MeV, 42 SEP events with protons accelerated to above 500~MeV have been identified during the last 20 years of the SOHO mission. The compiled event list (see Table \ref{tab:events}) has been compared to various other event lists including the GLE list.\newline 
For events with clear onset times the proton intensity spectra was derived in a time interval of two hours starting 30 minutes after the onset time. The spectral indices [$\gamma$] derived from the power law fit in the energy range below 800~MeV and the intensity at 500~MeV of the events have been compared (see Figure \ref{fig:fitparas} and Table \ref{tab:fit}).\newline
Based on this comparison, various results from the literature such as typical intensity increases above $2\cdot 10^{-3}$(cm$^2$ s sr MeV)$^{-1}$ at 500~MeV \citep{nitta} were validated. Furthermore, certain non-GLE and GLE events with surprisingly high and low intensity respectively were found in agreement with \citet{Thakur-2016}. \newline
Comparing the derived intensities at 500~MeV with the relative increase of the neutron monitor count rates during GLEs \citep{mccracken_2012}, a clear correlation has been found with exception of events with very small intensity at 500~MeV as well as GLE~69, which shows a particular large increase in neutron monitor count rate.

\acknowledgments
The SOHO/EPHIN project is supported under Grant 50~OC~1302 by the German Bundesministerium f\"ur Wirtschaft through the Deutsches Zentrum f\"ur Luft- und Raumfahrt (DLR).\newline
This project has received funding from the European Union's Horizon 2020 research and innovation programme under grant agreement No 637324.\newline
This work was carried out within the framework of the bilateral BMBF-NRF-project 'Astrohel' (01DG15009) funded by the Bundesministerium f\"ur Bildung und Forschung. The responsibility of the contents of this work is with the authors.\newline

\section*{Disclosure of Potential Conflicts of Interest}
The authors declare that they have no conflicts of interest.

\bibliographystyle{spr-mp-sola}
\bibliography{mybib}

\end{article} 
\end{document}